\documentclass{iopconfser}

\usepackage{a4wide,amssymb,cite,graphicx}
\usepackage{epsfig}
\usepackage{hyperref}

\begin{document}

\noindent Conference Proceedings for BCVSPIN 2024: Particle Physics and Cosmology in the Himalayas\\Kathmandu, Nepal, December 9-13, 2024 

\title{Exploring Physics beyond the Standard Model with Neutrinos}

\author{Rukmani Mohanta}
\affil{School of Physics, University of Hyderabad, Hyderabad - 500046, India}
\email{rmsp@uohyd.ac.in}

\begin{abstract}
Neutrinos, being elusive subatomic particles having only  weak interactions,   provide an ideal platform to look for physics beyond the Standard Model. In the present era of neutrino physics, various experiments are focusing towards the precision measurements of the oscillation parameters. Hence, various new physics scenarios which can affect the neutrino oscillation probabilities in matter, be in need of  careful investigation. There is a slight tension at the level of $2 \sigma$ between the recently reported results on the CP violating phase $\delta_{CP}$  by NOvA and T2K experiments, which can be attributed to the role of non-standard interactions (NSIs) in the propagation of neutrinos through matter. Another interesting possibility is that though neutrinos are electrically neutral, they can possess electromagnetic properties such as electric and magnetic dipole moments in the presence of new physics, and the neutrino oscillation experiments can put limits on them. In this work, we will focus on these two aspects and show how they can be used to probe the nature of new physics from model-building perspectives.
\end{abstract}
\section{Introduction}
The Standard Model (SM) of particle physics has been remarkably victorious in explaining most of the observed phenomena at the fundamental level, and its success is indubitable.
Despite its marvelous accomplishments, it fails to explain many cosmological observations such as matter dominance of the universe,  the nature and identity of the dark matter, dark energy, etc. Hence, the exploration of physics beyond the standard model is absolutely essential for understanding the true nature of our universe.  
The origin of small and non-zero neutrino masses inferred from various neutrino oscillation experiments remains so far as a mystery and thus provides an ideal platform for the exploration of physics beyond the Standard Model (BSM). Over the past few decades, there has been remarkable advancement in the precise determination of various neutrino oscillation parameters. However,  the CP violating phase $\delta_{CP}$ so far remains unascertained. The recent measurement on $\delta_{CP}$  by NOvA and T2K experiments, show slight disagreement at the level of $2 \sigma$, which may be considered as smoking-gun signal of non-standard interactions (NSIs) in the propagation of neutrinos through earth matter.
To realize the non-zero neutrino mass, many new ideas are proposed, and these new concepts are expected to have implications in many other sectors.  One such possibility amongst them is that neutrinos can possess electromagnetic properties like electric and magnetic dipole moments. Solar, accelerator and reactor experiments possibly could provide the direct measurement of magnetic moments and eventually put the  limits on them. 
\section{Scalar Nonstandard Interactions (NSIs)}
  
  Within the realm of BSM physics, the NSIs of neutrinos emerge as a notable and well-motivated phenomena. Various experiments have already delved into  NSI signals,  associated with  charge current (CC) as well as neutral current (NC)  interactions. It should be emphasized that the CC and NC NSIs are mediated through vector fields. The corresponding effects of NC-NSI appeared as a potential term in the Hamiltonian for neutrino oscillation and consequently affect the propagation of neutrinos between the source and detector. However, if the NSIs are mediated by a scalar field, the corresponding interactions contribute as a correction to the neutrino mass term as opposed to the matter potential. As a result, the effect of scalar NSI becomes independent of neutrino energy ($E_\nu$), while the vector-type NSI  scales linearly with $E_\nu$, leading to significantly different phenomenological consequences in neutrino oscillation experiments.

\begin{figure}[htb!]
    \centering
\includegraphics[scale=0.325]{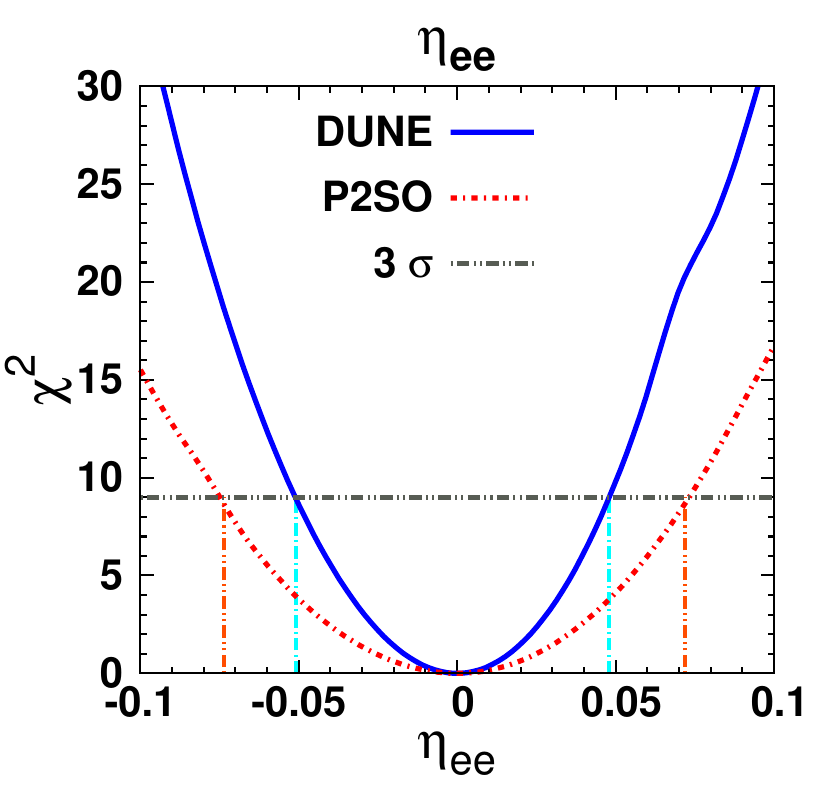}
\includegraphics[scale=0.325]{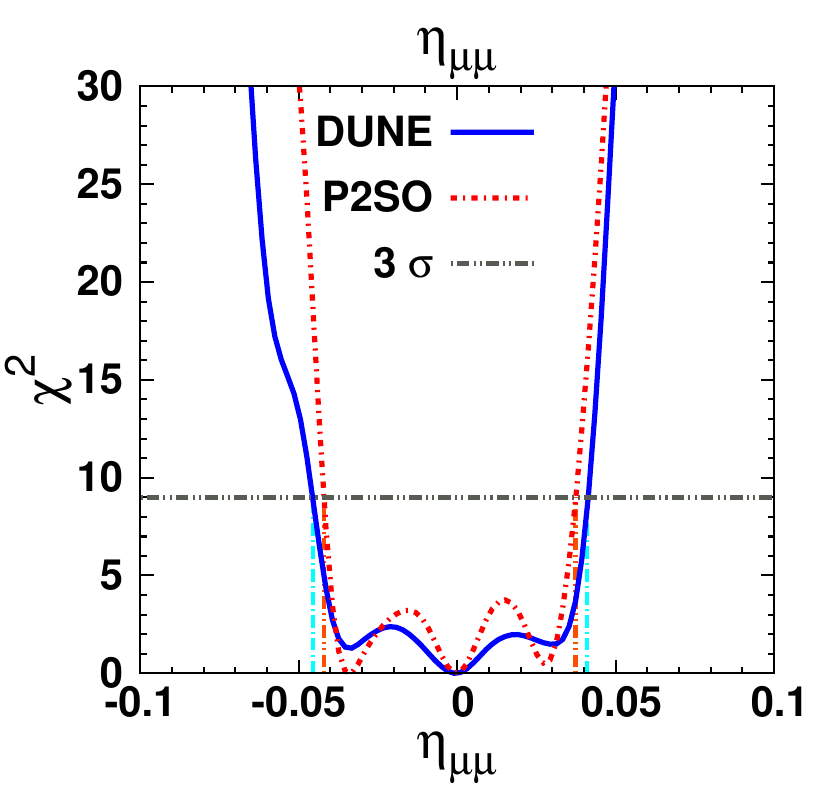}
\includegraphics[scale=0.325]{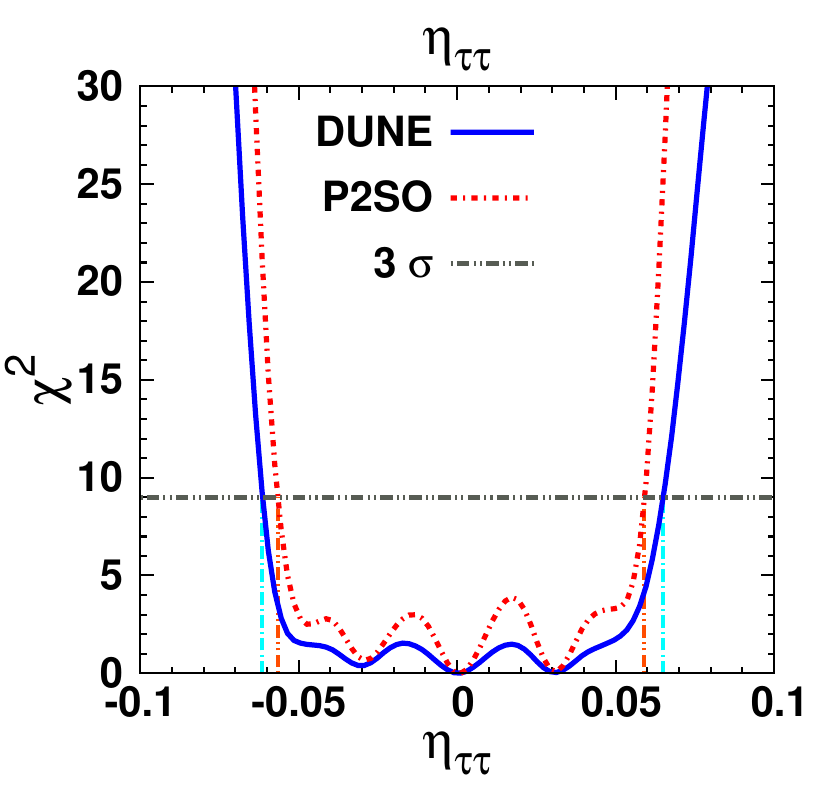}
\includegraphics[scale=0.325]{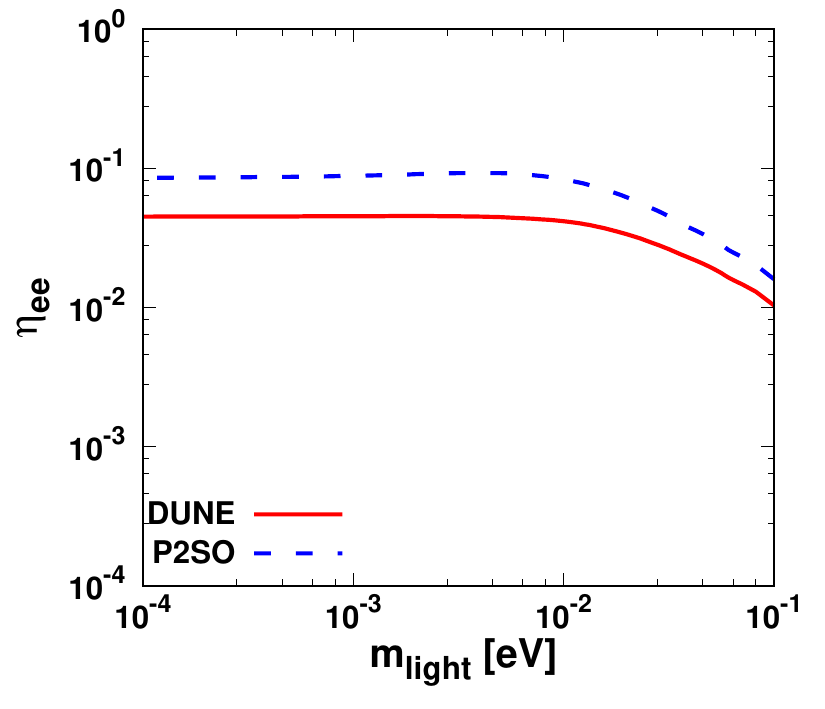}
    \includegraphics[scale=0.325]{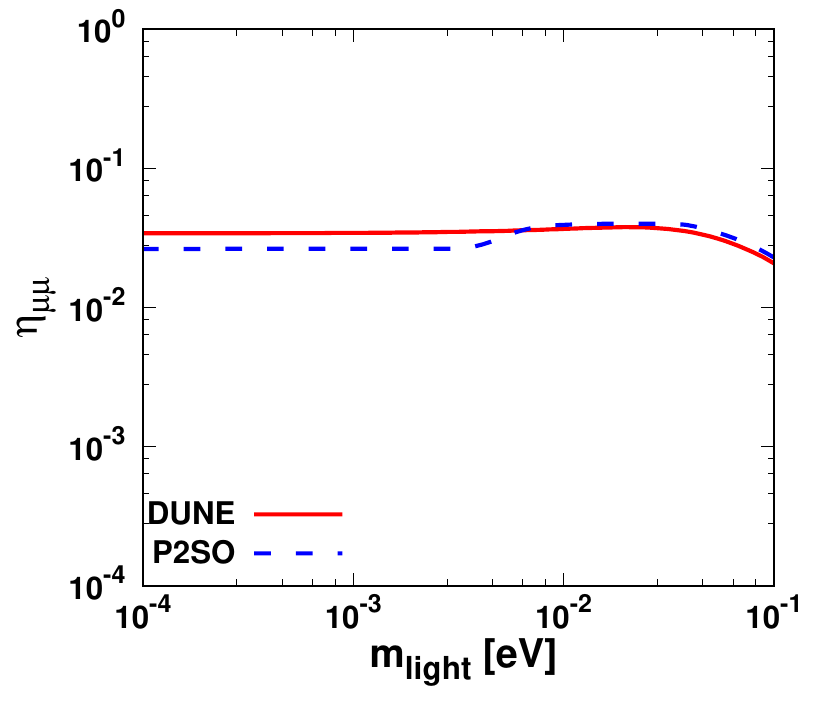}
    \includegraphics[scale=0.325]{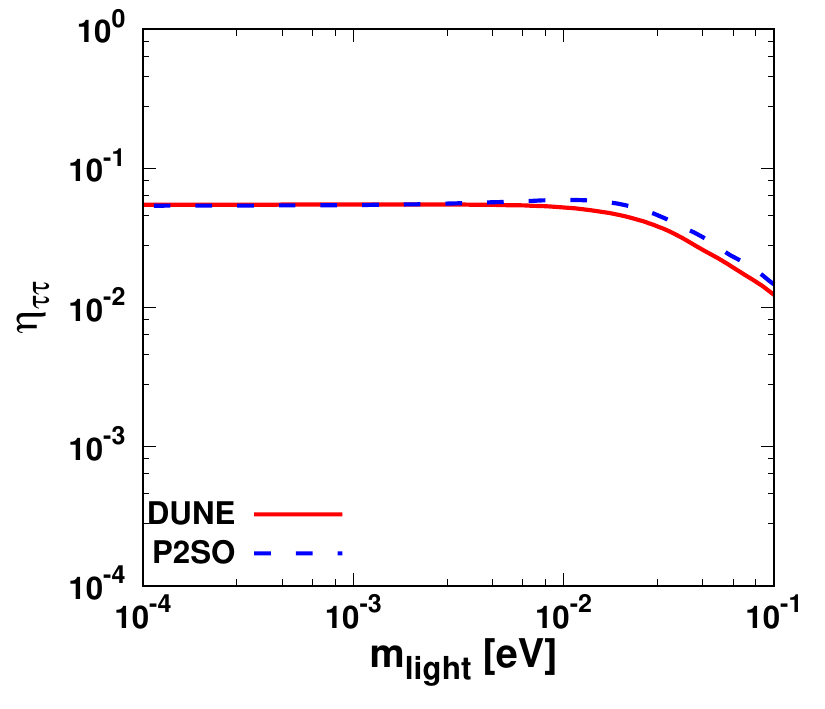}
    \caption{Bounds on the SNSI diagonal parameters ($\eta_{ee}, \eta_{\mu\mu}$ and $\eta_{\tau\tau}$) from DUNE and P2SO experiments (top panel).  Constraints on scalar NSI parameters in normal mass ordering (bottom panel) \cite{Ref-1}.}
    \label{fig-1}
\end{figure}

The effective Lagrangian describing the non-standard interaction between the neutrinos ($\nu$) and the fermions ($f$), mediated by a scalar field $\phi$  can be written as,
\begin{eqnarray}
 \mathcal{L}_{\rm eff} = \frac{y_f y_{\alpha \beta}}{m_\phi^2} (\overline{\nu}_\alpha\nu_\beta)(\bar{f}{f}),
 \label{lag}
\end{eqnarray}
where $y$’s represent the Yukawa couplings and $m_\phi$ denotes the mass of the scalar mediator. Thus, the modified Dirac equation in presence of SNSI can be expressed as,
\begin{eqnarray}
\overline{\nu}_\beta \left[i  \gamma^\mu \partial_\mu + \left(m_{\beta \alpha} + \frac{\sum_f N_f y_f y_{\alpha \beta}}{m_\phi^2}\right)\right] \nu_\alpha = 0\;,
\label{dir}
\end{eqnarray}
where $m_{\beta \alpha}$ stands for the Dirac mass matrix for the neutrinos and $N_f$ denotes the number density of fermion $f$. Thus, one can notice that the effect of SNSI manifests as a correction term to the neutrino mass matrix, which   can be parameterized as

\begin{eqnarray}
\delta m =  \sqrt{|\Delta m^2_{31}|}
\left(
\begin{array}{ccc}
\eta_{ee} & \eta_{e\mu} & \eta_{e\tau}\\
\eta_{\mu e} & \eta_{\mu \mu} & \eta_{\mu \tau} \\
\eta_{\tau e} & \eta_{\tau \mu} & \eta_{\tau\tau}
\label{eta}
\end{array}
\right )\;,
\end{eqnarray}
where we have rescaled the size of $\delta m$ relative to $\sqrt{|\Delta m^2_{31}|}$ to make the SNSI parameters,  i.e., $\eta$ dimensionless. Thus, comparing Eqs.~\ref{dir} and \ref{eta}, we can write
\begin{eqnarray}
    \eta_{\alpha \beta} = \frac{1}{m_\phi^2 \sqrt{|\Delta m^2_{31}|}} \sum_f N_f y_f y_{\alpha \beta}\;.
\label{eta_d}
\end{eqnarray}
Here $\Delta m^2_{31} = m_3^2 - m_1^2$ is the atmospheric mass square difference.  
Now let us illustrate how  $\delta m$ modifies the Hamiltonian of neutrino oscillation. The effective Hamiltonian  in presence of scalar NSI can be written as
\begin{eqnarray}
  H = E_\nu + \frac{\mathcal{M} \mathcal{M}^\dagger}{2E_\nu} + {\rm diag}(\sqrt{2}G_F N_e,0,0)\;.
  \label{ham}
\end{eqnarray}
 Here, the mass term $\mathcal{M}$ can be expressed as
\begin{eqnarray}
 \mathcal{M} &=& U~{\rm diag}(m_1, m_2, m_3)~U^\dagger + \delta m \;.
  \label{h} 
 \end{eqnarray}
 It is interesting to note that, in the presence of SNSI, the neutrinos oscillation probabilities will be dependent on the absolute neutrino mass $m_1$. In order to constrain the  SNSI parameters, we consider two upcoming long-baseline experiments: DUNE (Deep Underground Neutrino Experiment) and P2SO (Protvino to super-Orca). We simulate these two experiments using GLoBES software with modified probability engine \cite{Ref-1, Ref-2}. The results obtained are shown in the upper panel of  Figure \ref{fig-1}.  These plots are generated for normal ordering and taking the value of the lightest neutrino mass  $m_1 = 10^{-5}$ eV.  
From the figure, we see that stringent bounds can be obtained on these parameters, i.e.,  $\eta_{\alpha \alpha} \sim {\cal O}(10^{-2})$ from both the experiments. To see how the bounds on the diagonal SNSI parameters change with respect to the lowest neutrino mass $m_1$,  we have plotted in the lower panel of Fig.~\ref{fig-1}, the upper bound of the SNSI parameters at 3$\sigma$ as a function of $m_1$.  We can notice from the figure that for the lightest neutrino mass below the order $10^{-2}$ eV, the constraints on SNSI parameters are almost unchanged. 

\section{Electromagnetic properties of Neutrinos}

To demonstrate the neutrino mass,  magnetic moment and dark matter phenomenology in a single platform, we consider an extension of the SM with three additional vector-like fermion triplets $\Sigma_k$,  and two inert scalar doublets $\eta_j$ \cite{Ref-3}. In addition, we impose an additional $Z_2$ symmetry to illustrate neutrino phenomenology at one-loop level and also for the the dark matter stability. The particle content of the model along with their respective charges  are presented in Table. \ref{typeiii_model}.

\begin{table}[htb]
\begin{center}
\begin{tabular}{|c|c|c|c|c|}
	\hline
			& Field	& $ SU(3)_C \times SU(2)_L\times U(1)_Y$ & $Z_2$\\
	\hline
	\hline
	Leptons	& $\ell_{L} = (\nu, e)^T_L$	& $(\textbf{1},\textbf{2},~  -1/2)$ & $+$\\
			& $e_R$							& $(\textbf{1},\textbf{1},~  -1)$	& $+$\\
	&  $\Sigma_{k(L,R)}$ & $(\textbf{1},\textbf{3},0)$ & $-$ \\
\hline	
	Scalars	& $H$							& $(\textbf{1},\textbf{2},~ 1/2)$	&  $+$\\ 
					& $\eta_j$							& $(\textbf{1},\textbf{2},~ 1/2)$	& $-$\\
			\hline
	\hline
\end{tabular}
\caption{Particle content along with their charges in the present model.}
\label{typeiii_model}
\end{center}
\end{table}
 The relevant interaction Lagrangian involving the new particles  is given by
\begin{eqnarray}
\mathcal{L}_{\Sigma} = y^\prime_{\alpha k} \overline{\ell_{\alpha L}}\Sigma_{kR} \tilde{\eta}_{j} + y_{\alpha k} \overline{\ell^c_{\alpha L}}i\sigma_2 \Sigma_{kL} \eta_{j}  + \frac{i}{2}{\rm Tr}[\overline{\Sigma}\gamma^\mu D_\mu\Sigma] - \frac{1}{2}{\rm Tr}[\overline{\Sigma}M_\Sigma\Sigma] + {\rm h.c.}\;,
\end{eqnarray}
where  $\Sigma = (\Sigma_1,\Sigma_2,\Sigma_3)^T$. The  Lagrangian for the scalar sector is given as 
 \begin{eqnarray}
\mathcal{L}_{\rm scalar} = \left|\left(\partial_\mu  + \frac{i}{2} g~  \sigma^a W^a_\mu  + \frac{i}{2}g^\prime B_\mu\right) \eta_1\right|^2 +\left|\left(\partial_\mu  + \frac{i}{2} g~  \sigma^a W^a_\mu  + \frac{i}{2}g^\prime B_\mu\right) \eta_2\right|^2 - V(H,\eta_1, \eta_2),
\end{eqnarray}
where, the inert doublets are denoted by $\eta_j = \left (
		 \eta_j^+		,
		 \eta_j^0
	\right )^T$, with $\eta_j^0 = \displaystyle{({\eta^R_j+i\eta^I_j})/{\sqrt{2}}}$. $V(H, \eta_1, \eta_2) $ represents the scalar potential, from which the mass matrices of the charged and neural scalar components can be obtained, which upon diagonalization  yield the
 physical masses.

\begin{figure}
    \includegraphics[width=6.2cm,clip]{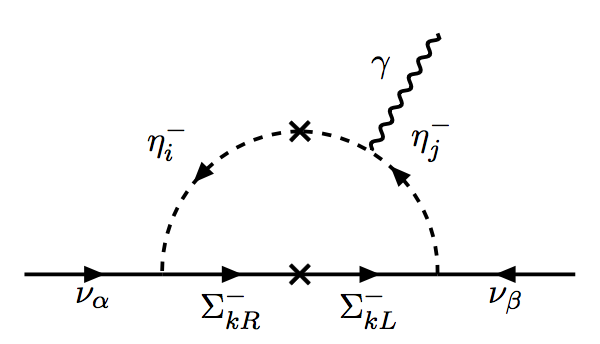}
\includegraphics[width=6.2cm,clip]{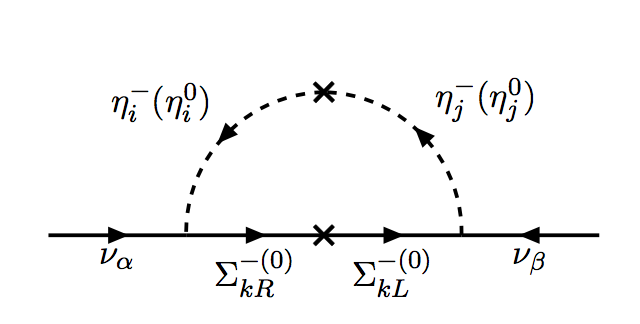}
    \caption{One-loop contribution for the transition magnetic moment (left panel) and light neutrino mass (right panel).}
    \label{numag_feyn}
\end{figure}

Although  neutrinos are electrically neutral, they can have an electromagnetic interaction at the loop level.
In this model, the transition magnetic moment arises from the one-loop diagram as shown in the left panel of Fig. \ref{numag_feyn}, and the corresponding contribution can be expressed as  \cite{Ref-3}
\begin{eqnarray}
(\mu_{\nu})_{\alpha \beta} = \sum^3_{k=1}\frac{({Y^2})_{\alpha \beta}} {8\pi^{2}}M_{\Sigma^+_k}   \bigg[\frac{(1+\sin 2\theta_C)}{M^{2}_{C2}} \left(\ln \left[\frac{M^{2}_{C2}}{M^2_{\Sigma^+_k}}\right]-1\right) 
+ \frac{(1-\sin 2\theta_C)}{M^{2}_{C1}} \left(\ln \left[\frac{M^{2}_{C1}}{M^2_{\Sigma^+_k}}\right]-1\right) \bigg],
\label{mag_eqn}
\end{eqnarray}
where $y = y^\prime = Y$ and $({Y^2})_{\alpha \beta}= {Y}_{\alpha k} {Y}_{k \beta}^T$.

The contribution for the neutrino mass  can be generated at one-loop level as shown in the right panel of  Fig. \ref{numag_feyn} and the corresponding contribution can be expressed as \cite{Ref-3}
\begin{eqnarray}
&&({\cal M}_{\nu})_{\alpha\beta} = \sum_{k=1}^3 \frac{({Y^2})_{\alpha \beta}}{32 \pi^{2}}M_{\Sigma^+_k}
 \Bigg[\frac{(1+\sin2\theta_C)M_{C2}^2}{M_{\Sigma^+_k}^{2}-M_{C2}^2}\ln\left(\frac{M^2_{\Sigma^+_k}}{M^2_{C2}}\right)+ \frac{(1-\sin2\theta_C)M_{C1}^2}{M_{\Sigma^+_k}^{2} -  M_{C1}^2}\ln\left(\frac{M^2_{\Sigma^+_k}}{M^2_{C1}}\right)  \Bigg]  \nonumber\\
&&~~~~+\sum_{k=1}^3 \frac{{(Y^2)}_{\alpha \beta}}{32 \pi^{2}} M_{\Sigma^0_k}
 \Bigg[\frac{(1+\sin2\theta_R)M_{R2}^2}{M_{\Sigma^0_k}^{2}-M_{R2}^2}\ln\left(\frac{M^2_{\Sigma^0_k}}{M^2_{R2}}\right) + \frac{(1-\sin2\theta_R) M_{R1}^2}{M_{\Sigma^0_k}^{2} -  M_{R1}^2}\ln\left(\frac{M^2_{\Sigma^0_k}}{M^2_{R1}}\right)  \Bigg]  \nonumber\\
&&~~~~-\sum_{k=1}^3 \frac{{(Y^2})_{\alpha \beta}}{32 \pi^{2}} M_{\Sigma^0_k}
 \Bigg[\frac{(1+\sin2\theta_I)M_{I2}^2}{M_{\Sigma^0_k}^{2}-M_{I2}^2}\ln\left(\frac{M^2_{\Sigma^0_k}}{M^2_{I2}}\right)+ \frac{(1-\sin2\theta_I)M_{I1}^2}{M_{\Sigma^0_k}^{2} -  M_{I1}^2}\ln\left(\frac{M^2_{\Sigma^0_k}}{M^2_{I1}}\right)  \Bigg].
\label{mass_eqn}
\end{eqnarray}
\section{Dark Matter phenomenology}

Considering the new scalar particles as dark matter candidates, here we investigate their phenomenology. The dark matter relic density receives contributions from all the inert scalar components through their annihilation and co-annihilation processes.  The abundance of dark matter relic density can be computed using the standard formula,
\begin{equation}
\label{eq:relicdensity}
\Omega h^2 = \frac{1.07 \times 10^{9} ~{\rm{GeV}}^{-1}}{  M_{\rm{Pl}}\; {g_\ast}^{1/2}}\frac{1}{J(x_f)}\;,
\end{equation}
where, $M_{\rm{Pl}}=1.22 \times 10^{19} ~\rm{GeV}$ is the Planck mass and $g_\ast = 106.75$ represents the  total number of effective relativistic degrees of freedom, while the function $J(x_f)$ is related to the thermal-averaged  annihilation cross section of WIMP pair. 

The  direct search signals mainly come from the scattering off the scalar dark matter from the nucleus via the Higgs boson.  Thus, the DM-nucleon cross section in Higgs portal can provide a spin-independent (SI) cross section, whose sensitivity can be checked with stringent upper bound of LZ-ZEPLIN experiment. 
The corresponding cross section is given as 
\begin{equation}
    \sigma_{\rm SI} = \frac{1}{4\pi}\left(\frac{M_n M_{R1}}{M_n + M_{R1}}\right)^2 \left(\frac{\lambda_{L1} \cos^2\theta_R + \lambda_{L2}\sin^2\theta_R}{2 M_{R1} M_h^2}\right)^2 f^2 M_n^2,
\end{equation}
where, $M_n$ denotes the nucleon mass, nucleonic matrix element $f \sim 0.3$. We have used micrOMEGAs to compute relic density and also DM-nucleon cross section. 
\section{Analysis}
\begin{figure}[htb!]
\centering
\includegraphics[width=5.75 cm, clip]{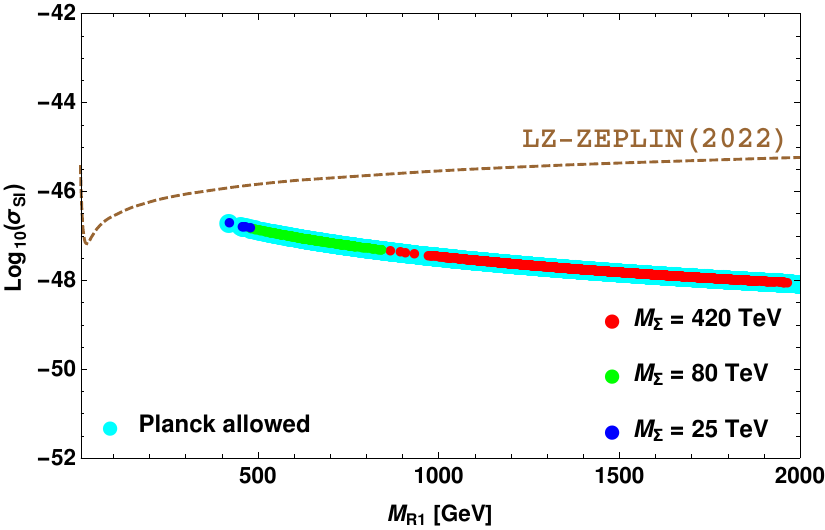}
\hspace{0.25 cm}
\includegraphics[width=6.53 cm, clip]{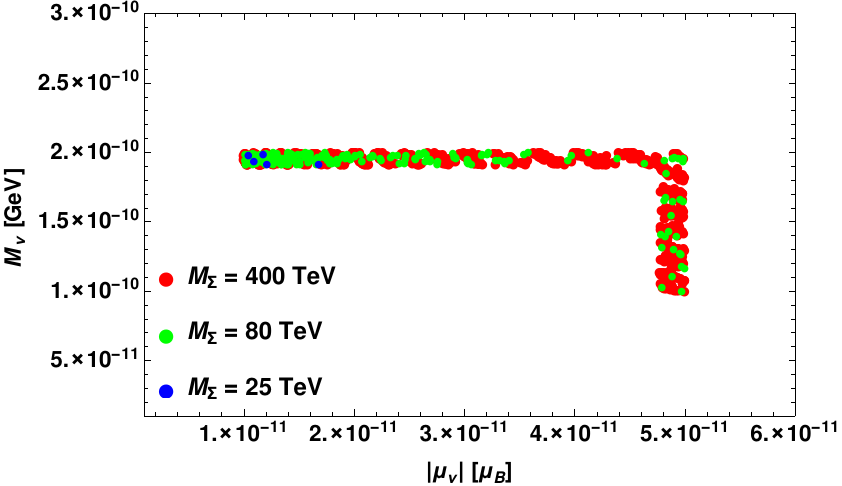}
\caption{The SI WIMP-nucleon cross section is projected  as a function $M_{R1}$, with dashed brown line of LZ-ZEPLIN upper limit \cite{LZ} (left panel). Cyan data points satisfy Planck limit \cite{Aghanim:2018eyx}. Blue, green and red data points satisfy neutrino mass and magnetic moment for a specific set of values for fermion triplet and Yukawa (right panel).}
 \label{scatter1}
\end{figure}

\begin{figure}[htb!]
\centering
\includegraphics[scale=0.585]{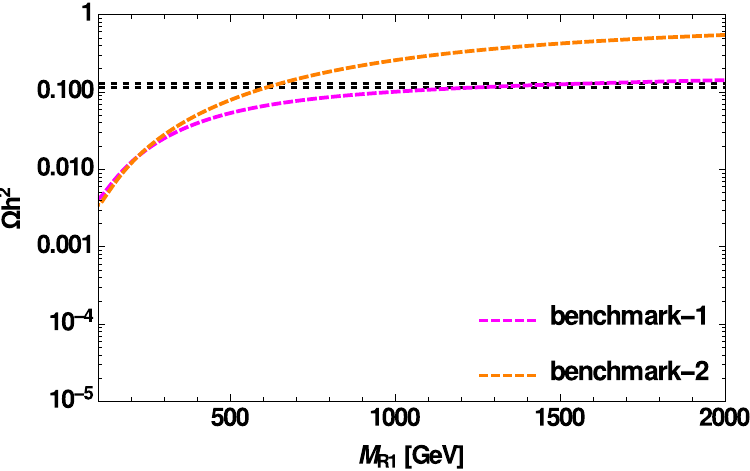}
\includegraphics[scale=0.425]{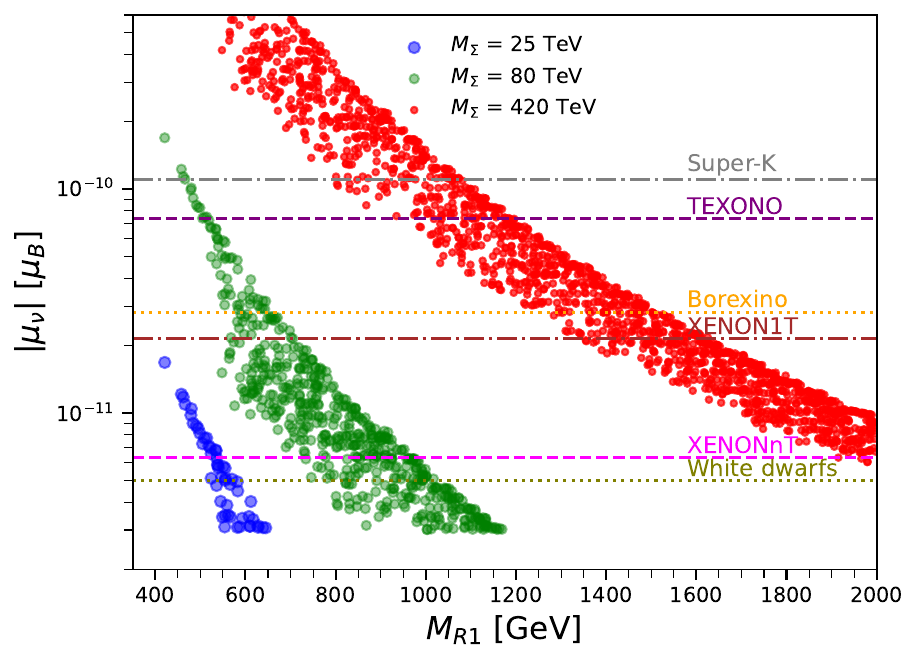}
 \caption{Left panel depicts the relic density as a function of dark matter mass for the specific benchmark values from table-\ref{tab_benchmark}. Right panel portrays the allowed region of neutrino magnetic moment with the  dark matter mass. Horizontal colored lines represent the upper bounds from different experiments.}
 \label{relic_plot}
\end{figure}

We consider the lightest inert scalar as $\phi^R_1$ with mass $M_{R1}$ and the masses of the other scalar particles are related to $M_{R1}$ through the mass splittings: $\delta$, $\delta_{\rm IR}$ and $\delta_{\rm CR}$. Thus, the masses of the rest of the inert scalars can be obtained using the relations:
\begin{eqnarray}
 M_{R2} - M_{R1} = M_{I2} - M_{I1} = M_{C2} - M_{C1} = \delta,~~  M_{Ri} - M_{Ii} = \delta_{\rm IR}, \quad M_{Ri} - M_{Ci} = \delta_{\rm CR}\;,
\end{eqnarray}
where, $i = 1,2$.
Performing a scan over model parameters as given below, we obtained the allowed parameter space, consistent with experimental bounds associated with both dark matter as well as neutrino sectors: 
\begin{eqnarray}
100 ~{\rm GeV} \le M_{R1} \le 2000 ~{\rm GeV}, ~~~
0.1 ~{\rm GeV} \le \delta < 200 ~{\rm GeV}, \quad 0.1 ~{\rm GeV} \le \delta_{\rm IR}, \delta_{\rm CR} \le 20 ~{\rm GeV}.
\end{eqnarray}
First, we eliminate the parameter space by imposing Planck constraint on relic density \cite{Aghanim:2018eyx} and then compute the  spin-independent DM-nucleon cross section for the available parameter space. In the left panel of Fig. \ref{scatter1}, we project the cross section as a function of $M_{R1}$  with cyan data points,  where the dashed brown line represents the LZ-ZEPLIN upper limit \cite{LZ}. Choosing a representative set of values for the  fermion triplet mass and  Yukawa couplings, with the obtained parameter space, one can satisfy the  aspects of neutrino mass and mixing phenomenology. Considering the triplet masses as $25, 80$ and $420$ TeV along with  suitable Yukawa, represented by  blue, green and red data points, satisfy the neutrino magnetic moment and light neutrino mass in the desired range simultaneously, as projected in the right panel. We observe that a wide region of dark matter mass is favored as we move towards high scale (triplet mass). 
%
Using two specific benchmark values (shown as BM-1 and BM-2 in table \ref{tab_benchmark})  which are favourable for explaining both neutrino and dark matter prospects, we project relic abundance of scalar dark matter in the left panel of  Fig. \ref{relic_plot}. In the right panel of Fig. \ref{relic_plot}, we display neutrino magnetic moment as a function of dark matter mass, by considering specific set of values assigned to the triplet fermion mass. It is clear from the plots that the model parameters are able to provide neutrino magnetic moment in the range $10^{-12}\mu_B$ to $10^{-10}\mu_B$, sensitive to the upper limits from various experiments, such as Super-K, TEXONO, Borexino, XENON1T, XENONnT  and white dwarfs (colored horizontal lines). Thus, from all the above discussions, it is evident that the proposed model can provide a well-consistent framework for a correlative study of neutrino mass, magnetic moment mass and dark matter phenomenology.
\begin{table}[htb]

\begin{center}
\begin{tabular}{|c|c|c|c|c|c|c|c|c|c||c|c|}
	\hline
			& $M_{R1}$ [GeV]	& $\delta$ [GeV] & $\delta_{\rm CR}$ [GeV] & $\delta_{\rm IR}$ [GeV] & $M_{\Sigma}$ [TeV]& $|\mu_\nu|$ [$\mu_B$] & $\Omega {\rm h}^2$ \\
	\hline
	BM-1 & $1472$	& $101.69$	& $9.03$ & $0.35$ & $420$ & $2.73\times 10^{-11}$ & $0.123$ \\
	\hline
	BM-2 & $628$	& $36.40$	& $4.38$ & $3.45$ & $80$ & $3.03\times 10^{-11}$& $0.119$\\
	\hline
\end{tabular}
\caption{Set of benchmark points from the consistent parameter space.}
\label{tab_benchmark}
\end{center}
\end{table}
%

\section{Conclusions}
In this work, we have demonstrated the exploration of Physics beyond the Standard Model with neutrinos. First, we discussed the effects of scalar non-standard interactions on neutrino oscillations and have shown that the upcoming  long-baseline experiments DUNE and P2SO can put the constraints on the SNSI parameters as $\eta_{\alpha \alpha} \sim {\cal O}(10^{-2})$.  Secondly, we present a discussion on electromagnetic properties of neutrinos and   tried to address neutrino mass, magnetic moment and dark matter phenomenology in a common platform. For this purpose, we  extended the SM with three vector-like fermion triplets and two inert scalar doublets to realize neutrino mass generation through Type-III radiative scenario.  The pair of charged scalars help in generating neutrino magnetic moment while all the charged and neutral scalars play the role for obtaining light neutrino mass at one-loop level. Additionally, all the inert scalars involved through annihilation and co-annihilation channels to provide total dark matter relic density, consistent with Planck  data and also provide a suitable cross section with nucleon, sensitive to LZ-ZEPLIN upper limit. Finally, we have also demonstrated  that the model is able to provide neutrino magnetic moment in a wide range ($10^{-12}\mu_B$ to $10^{-10}\mu_B$), in the same ball park of Borexino, Super-K, TEXONO, XENONnT and white dwarfs. 

{\bf Acknowledgments}

RM would like to thank University of Hyderabad IoE project grant no. RC1-20-012.

\end{document}